\documentstyle[preprint,aps,epsf]{revtex}
\tightenlines                  
\global\let\epsfloaded=Y 
\begin{document}
\pagestyle{empty}                                      
\preprint{
\font\fortssbx=cmssbx10 scaled \magstep2
\hbox to \hsize{
\hfill $
\vtop{
 \hbox{ }}$
}
}
\draft
\vfill
\title{Incompressible Hydrodynamics on a Noncommutative D3-Brane}
\vfill
\author{Tzi-Hong Chiueh}
\address{
\rm Department of Physics, National Taiwan University,
Taipei, Taiwan 10764, R.O.C.}

%
%
\vfill         
\maketitle      
\begin{abstract}  
In the Seiberg-Witten limit, the low-energy dynamics of N weakly 
coupled identical open strings on a D3-brane can behave as 
two-dimensional incompressible hydrodynamics.  Classical vortices 
are frozen in the fluid and described by an action expressed in 
terms of two canonical conjugate fields, which can be taken as 
the new space coordinate. The noncommutative space naturally arises 
when this pair of conjugate fields are quantized.  To the lowest 
order of $\hbar$, the vorticity can replace the background $B$-field 
on the D3 brane, thereby yielding a spatially and temporally varying
noncommutative parameter $\theta^{ij}$. Demanding a quantum
area-preserving transformation between two classical inertial-frame
coordinates, we identify the classical solitons that survive in the
noncommutative space, and they turn out to be the "electric field" 
solutions of the Dirac-Born-Infeld Lagrangian created by a
$\delta$-function source.  The strongly magnetized electron gas in a
semiconductor of finite thickness is taken as a case study, where 
similar quantum column vortices as those in a D3 brane can be present.  
The electric charges contained in these electron-gas column vortices 
are quantized, but in a way different from those in the sheet vortices
that produce the fractional quantum Hall effect.    
 
\end{abstract}
%
%
\pacs{PACS numbers:
 }
%
%
\pagestyle{plain}

\section{Introduction}

Gravity should be described by quantum
mechanics at the Planck epoch.  In seeking
the quantum theory of gravity, 
it requires the spacetime to be subject to
quantum fluctuations.  Thus, the spacetime becomes fuzzy over some
length and time scales, where the notion of classical spacetime breaks
down.  It was first proposed in 1986 that the string theory can naturally
give rise to a fuzzy spacetime geometry\cite{1}.  The matrix theory
compactified on a torus in the presence of a background field provides a 
concrete example of the noncommutative space\cite{2}.  On a different
path, it has also been shown that the string on a D-brane world volume 
can also yield a noncommutative spacetime\cite{3}.  The fuzzy space found 
on the D-brane has recently become the focus of attention because it
does not rely on the space to be compactified into a torus. 

Open strings have endpoints, which anchor on the D-branes.  
Such open strings have interesting connections with 
the Yang-Mills theory when there
exists an antisymmetric tensor field $B_{\mu\nu}$
\cite{2,3,4,5,6,7,8,9}.
Normally, the antisymmetric field is magnetic-like; 
that is, the string 
endpoints do not experience the electric field $E_i=B_{0i}$.
Moreover, due to the presence of the background 
field on the D-brane, the dynamics
of the string endpoints is different from that of 
the string interior.
The disparity of dynamics can be built in their kinematics
by different metrics\cite{9,10}.  
The closed-string metric
$g_{\mu\nu}$ describes that of the original space.
The open-string metric $G_{\mu\nu}$ is an effective one,
describing the oscillating motion of string endpoint on the D-brane
due to the coupling to the oscillating string bulk. 

The effective metric $G_{\mu\nu}$ is related to 
$g_{\mu\nu}$ via the field $B_{\mu\nu}$ by the relation
\begin{equation}
G_{\mu\nu}=g_{\mu\nu}-(\alpha')^2 B_{\mu\xi}(g^{-1})^{\xi\eta}B_{\eta\nu},
\end{equation}
where $(\alpha')^{-1}$ is the string tension,
and raising or lowering indices always refers to the original
metric $g_{\mu\nu}$ and the indices run from $0$ to $p$ for a 
$p$-dimensional brane, the Dp-brane.
Equation (1) shows that only the components
of $B_{\mu\nu}$ that are parallel to 
the D-brane matter, and hence other
components can be set to zero.

The following simple picture, which is generalized to include an
"electric" field, can be helpful for understanding 
the basic dynamics of an open string on a flat D2-brane.  
For notational simplicity, we let 
$B\equiv B_{12}$, to be conceived
as the magnetic field, and $E_x\equiv B_{01}$, $E_y\equiv B_{02}$
as the electric field.  The string bulk does not interact with
a uniform background field, and only the endpoint does.
We thus let a unit charge be attached to the string endpoint, 
which moves on the $x-y$ plane perpendicular to
the magnetic field.  Apart from the electromagnetic forces, 
the endpoint also experiences
a tension force, ${\bf T}(x,y)$, parallel to the D2-brane,  
as the string is generally not perpendicular to the brane.
The string endpoint has a zero mass, and the combined electric
and tension forces,
${\bf F}={\bf E}+{\bf T}(x,y)$, must instantaneously be 
balanced by a magnetic force $q\dot{\bf x}\times B\hat z$.  
It follows that the endpoint velocity
$\dot{\bf x}={\bf F}\times\hat z/B$.  As the velocity is always
perpendicular to the force, there is no energy exchange
between the string endpoint and the electromagnetic/tension fields. 
The origin of this peculiar feature is that the string endpoint
has zero inertia, thus containing no energy.  

In mathematical terms, 
the endpoint equation of motion reads:
\begin{equation}
B_{\mu\nu}\partial_t x^\nu=-(\alpha')^{-1}g_{\mu\nu}\partial_\sigma x^\nu,
\end{equation}
where $\sigma$ is the coordinate 
along the string and $t$ is the time.  The space component of (2) shows
that the electromagnetic
forces are on the left and the tension force is on the right.
In the non-relativistic limit, the time component of (2) indicates
that the velocity is perpendicular to the electric field, i.e., 
${\bf v}\cdot {\bf E}=0$, meaning that the tension force is always
parallel
to the electric field, ${\bf T}\parallel {\bf E}$.  Since the tension
force $T_i=-\alpha'\partial_\sigma x_i$, we may express it as 
$-\partial_i(\alpha'\partial_\sigma r^2/2)$, as long as 
$\partial_\sigma$ and $\partial_i$ commute which is usually the case.
We may also choose the electric field to also be expressed as
$E_i=-\partial_i\phi$, and the combined force
$F_i=-\partial_i[\phi+\alpha'\partial_\sigma(r^2/2)]\equiv
-\partial_i\bar\phi$.  The velocity becomes
$\dot{\bf x}=-\nabla\times(\hat z\bar\phi/B)$ for a uniform 
$B$-field.  This gives rise to an important kinematic constraint that 
the endpoint motion is incompressible, $\partial_i\dot x^i=0$.
 
Apart from the above characteristics, the endpoint is also subject to
oscillations in response to
the oscillation modes within the string bulk.
A Green's function calculation\cite{11,12,13}
shows that the 
auto-correlation function of the endpoint motion reads
\begin{equation}
\langle x^\mu(t) x^\nu(0)\rangle=
i\alpha'(G^{-1})^{\mu\nu}\log t^2+(i/2)\theta^{\mu\nu}\epsilon(t),
\end{equation}
where $\epsilon(t)$ equals $1$ or $-1$ for positive 
or negative time $t$, and
the anti-symmetric tensor $\theta^{\mu\nu}$ is given as
\begin{equation}
\theta^{\mu\nu}\equiv 2\pi(\alpha')^2(g^{-1} B G^{-1})^{\mu\nu}.
\end{equation}
The first term of the auto-correlation function represents 
the time symmetric coupling to the bulk oscillations where
the open-string metrics $G_{\mu\nu}$ plays the role of an
effective metric; the second term carries the time direction,
representing a net displacement around a circle
acquired from the secular drift (zero-frequency mode) of endpoint motion.
The above Green's function is obtained in the frame where
the "electric" field $B_{0i}$ is absent.

When the string endpoint is attached to a D3-brane, 
a situation relevant to the present work, 
one can think of the charged particle to be in a three-dimensional 
(3D) space in the presence of a magnetic field lying within
the 3D brane world volume.  The magnetic field lines
break the isotropy of the 3D space into two transverse 
directions and one longitudinal direction.  
The string extends out of the 3D space into the higher
dimensions, with its tension force oriented to an
arbitrary direction in the 3D space.  
In the direction perpendicular to the magnetic
field line, the force balance is identical to that on
a D2-brane described above.  But the
component of tension force along the magnetic field line is now 
unbalanced, causing the endpoint to oscillate freely along the field line
in response to the bulk oscillations.

On a D3-brane,
the Seiberg-Witten map\cite{9} 
takes the decoupling limit, $g_{00}=-g_{33}=-1$, 
$g_{ij}\to\epsilon^2\delta_{ij}$
and $\alpha'\to\epsilon$, with $\epsilon\to 0$ and
$i,j$ running from $1$ to $2$.
A straightforward calculation shows that
\begin{equation}
\theta^{ij}\to 2\pi(B^{-1})^{ij}=2\pi B^{-1}\epsilon^{ij},
\end{equation} 
and $\theta^{\mu\nu}\to 0$, otherwise.  In the same limit,
we also have
\begin{equation}
G_{ij}\to {\alpha'^2 B^2\over\epsilon^2}\delta_{ij}, \ \ \ 
G_{00}=-1+{\alpha'^2 B^2 v_jv^j\over\epsilon^2 c^2}, \ \ \ 
G_{0i}=G_{i0}={\alpha'^2 B^2 v^i\over\epsilon^2 c},
\end{equation}
where ${\bf v}/c \to {\bf E}\times \hat z/B$, the endpoint velocity 
given in (2) in this decoupling limit. 
The open-string metric $G_{\mu\nu}$ and
$\theta^{ij}$ are both finite.
The string now becomes a rigid wire, where the string oscillations
are hard to excite, according to (3), and to influence the endpoint
motion.  The net displacement around a circle in (3) 
remains finite in this limit.

In the presence
of a constant electric field ${\bf E}$, the endpoint moves at a 
constant velocity, and this electric field can be transformed away by
choosing an appropriate reference frame.
It is in this frame that the net displacement of endpoint motion in (3)
is calculated, and the quantum field theory on the D-brane
becomes such that the D-brane coordinates are noncommutative.
That is, the coordinates on the brane
are operators that satisfy\cite{3}
\begin{equation}
[x^i, x^j]=i\theta^{ij}.
\end{equation}
Physically, what happens here is that the momentum, 
$p_j=-i\hbar\partial_j+eA_j$, 
is dominated by the "diamagnetic"
current, $eA_j(x)$, in the presence of a strong field.  
By the relation, 
$[eA_j(x),x_k]\to [p_j,x_k]=-i\hbar\delta_{jk}$, 
the open-string space becomes noncommutative.  

In view of the complications arising from 
the operator algebras,
one may alternatively represent functions of 
noncommutative coordinates 
by functions of commutative coordinates, except that 
the product of any two of the ordinary functions, 
e.g., $f$ and $h$, 
is replaced by the star-product\cite{14}, defined as
\begin{eqnarray}
f*h=&&fh+{i\over 2}\theta^{ij}(\partial_i f)(\partial_j h)
-{1\over 8}\theta^{kl}\theta^{ij}(\partial_k\partial_i f)
(\partial_l\partial_j h)\nonumber\\
&&-{1\over 12}\theta^{kl}\partial_k\theta^{ij}
((\partial_i f)(\partial_l\partial_j h)-
(\partial_l\partial_i f)(\partial_j h))
+ O(\theta^3),
\end{eqnarray}
where all products on the right are ordinary products.
The commutator now is $[f,h]\equiv f*h-h*f$. 
Substituting $x_i$ and $x_j$ for $f$ and $h$, one recovers (7).  
It is straightforward to show, by the symmetry of indices, that
$[f,h]=i\theta^{ij}(\partial_i f)(\partial_j h)+O(\theta^3)$, 
where the contribution from the second-order, $O(\theta^2)$, 
terms vanishes.

Note that the coefficient, $1/12$, of the term with 
$\partial_k\theta^{ij}$ in (8)
is chosen such that the star product, to the order 
$O(\theta^2)$, obeys
the associativity\cite{14}, $(f*q)*h=f*(q*h)$.  Apart from
an appropriate coefficient, the associativity 
further requires  
\begin{equation}
H_{ijk}\equiv \partial_i B_{jk}+\partial_j B_{ki}
+\partial_k B_{ij}=0,
\end{equation}
or $\nabla\cdot {\bf B}=0$ on the D3-brane.
Apparently, a magnetic-like field defined as
$B_{ij}\equiv\partial_i A_j-\partial_j A_i $
does satisfy (5).  The antisymmetric field
$B_{\mu\nu}$ normally is a linear combination of the Neveu-Schwarz (NS)
field and a $U(1)$ gauge field.  In the present work, we shall 
assume that the
NS field dominantly contributes to $B_{\mu\nu}$.

In the reference frame void of the electric-like
NS field, the energy density of the NS field 
is proportional to $H_{ijk}H^{ijk}$ and hence any NS field
that makes $H_{ijk}=0$ contains no energy.  In a general
reference frame, the time component should be included 
and a vacuum NS field obeys
\begin{equation}
H_{\mu\nu\eta}\equiv\partial_\mu B_{\nu\eta}+\partial_\nu B_{\eta\mu}
+\partial_\eta B_{\mu\nu}=0,
\end{equation}
where the indices include $0$ and
the additional constraint from (10) on the D3-brane 
is the Faraday's law.  Again, as the energy denisty of the NS field 
is proportional to $H_{\mu\nu\eta}H^{\mu\nu\eta}$, the field energy 
is zero.

Equation (10) implies that
$B(x,y,t)$ is not only non-uniform but can 
also be time-dependent.  However,
the solution to (10) is not well-defined, since we
do not know what $B_{0i}$ should be.  
In light of the discussions made prior to (7),
we demand that the "electric"
field $B_{0i}$ vanishes in the rest
frame of the string endpoint, i.e., $B_{0i}'=0$.  It provides a
condition to relate the "electric" field to the endpoint
motion by $E_i\equiv B_{0i}=B_{ij}v^j/c$. 
This is the perfect conductor limit, which
permits the endpoint to move
across the "magnetic" field ${\bf B}$.  
The NS field so constructed remains a vacuum field and 
contain no energy; such an NS field differs from 
the $U(1)$ gauge field since
it lacks the other half of the Maxwell equations.  
In this paper, we shall
identify $B_{12}$ to be the vorticity of the incompressible fluid.
Fluid motion contains the kinetic energy of particles,
but the presence of vorticity does not add anything extra.

The above relation for $v_j$ and $E_i$ is still
somewhat ambiguous in the sense that one may regard either 
the velocity to be
generated by a given "electric" field or the "electric" field 
generated by a given velocity.  In the context of open strings,
we may adopt either view.  When there exist many open strings on the 
same D-brane, strings can interact.  One situation where such an interaction
is inevitable is when the two endpoints of a string are on the same
D-brane; the small tension force given by (2) couples them.
On the other hand, charged string endpoints can also create a $U(1)$ 
Coulomb electric field so as to couple to each other; the energy of
such a Coulomb field can be made negligibly small compared with 
the kinetic
energy when the denisty of endpoints is sufficiently large, as will be
addressed in more details in Sec.(IV).

Indeed it is well known that the guiding-center motion
of a strongly magnetized 3D dense electron gas is incompressible
and involves only 2D displacements\cite{15}.  Their motion is advanced
by a self-induced Coulomb electric field.
Such an electron-gas system, with an neutralizing background, can be the
classical counterpart of N weakly coupled open strings
on a D3-brane with noncommutative
coordinates.  We are so motivated to focus on the
D3-brane, with an emphasis on
the non-uniform $\theta^{ij}$ in the decoupling limit.  
We are also seeking collective effects that may occur due to 
the interactions among many open strings attached on the same
brane.

Since $B^k (\epsilon^{kij}B_{ij})$
is along one direction on the D3-brane, 
we let it be ${\bf B}=B\hat z$.  
The dynamics
of string endpoint in the $z$ direction is almost a free motion 
in the decoupling limit subject only to the open-string interactions, 
and in this limit 
the dynamics in $z$ direction is decoupled from that
perpendicular to ${\bf B}$, provided that
the interactions of open strings are relatively weak. 
Thus, the relevant dynamics on a D3-brane in the 
decoupling limit is essentially {\it two dimensional}.  
We further conceive that the two endpoints of the rigid
string anchor onto two neighboring parallel D3-branes, 
and that the NS fields
${\bf B}$ and ${\bf E}$ are inhomogeneous only in the 
two relevant space coordinates $x$ and $y$. The inhomogeneity
$\Delta B(\equiv |B-B_0|)\ll B_0$, where $B_0$ is the time-independent, 
space-averaged $B$.  The presence of a fairly uniform $B$
imposes a kinematic constraint on the two-dimensional 
endpoint motion, such that
it is incompressible, as discussed below (2).

This paper is organized as follows.  
Sec.(II) advances the connection of the two-dimensional 
incompressible fluid to the noncommutative space, where  
the action of
incompressible hydrodynamics is obtained, thereby
allowing one to identify the suitable
canonical coordinate for dealing with space quantization 
in the presence of a non-uniform and time-dependent $\theta^{ij}$.
Moreover, $(\theta^{ij})^{-1}$ is shown to be the vorticity, indicative
of that $\Delta B$ should replace $B$ in (5).     
We then make use of
this new coordinate to examine canonical transformations in Sec.(III) 
and obtain some special classical soliton solutions, 
which survive in the noncommutative space and which
turn out to be the solutions to
the Dirac-Born-Infeld Lagrangian.  In Sec.(IV), the strongly 
magnetized electron-gas column is studied as an example of quantum
incompressible hydrodynamics.  The quantized charge contained 
in the vortex is quantitatively predicted.
Discussions and conclusions are given in Sec.(V).

For clarity of presentation, we shall focus on the low-energy
physics, where the non-relativistic theory 
applies and the incompressibility
condition, central to the present discussions, 
has no ambiguity.  (In fact, the decoupling limit also
leads to the non-relativistic limit for the endpoint motion.)  Only the
flat D3-brane, where $g_{00}=-g_{zz}=-1$
and $g_{ij}=\epsilon^2\delta_{ij}$ with $i,j$ being the indices 
for either $x$ or $y$ coordinates, will
be considered.  

\section{Action for Incompressible Hydrodynamics}

In the conventional notion of a uniform $\theta^{ij}$, 
one may conceive the commutation relation (7) to be analogous
to that of $[x_i,p_j]=i\hbar\delta_{ij}$.  
The latter quantizes the two-dimensional phase space
into many cells of arbitrary shape but of same area. 
The noncommutative coordinates also quantize
the space perpendicular to ${\bf B}$
into many unit cells of arbitrary shape.  The quantization 
of phase space is possible only because the phase space in
classical mechanics is incompressible.  Similarly, 
the classical fields that precede
the quantum fields of
noncommutative geometry should also involve 
incompressible motion in
directions perpendicular to ${\bf B}$.
In a D3-brane, such classical motion of
identical particles or identical string endpoints satisfies 
$\nabla_\perp\cdot{\bf v}$=0, where $\perp$ refers to 
the direction perpendicular to ${\bf B}$.

When $\theta(\equiv\theta^{12})$ is non-uniform and time-dependent, 
the quantized two-dimensional spatial cells on the D3-brane
will have different sizes at different locations and time.  
In contrast to the phase-space quantization, 
a spatio-temporal varying 
$\theta$ seems to make the quantization of spatial
cells at odds.  It is therefore instructive to 
again examine how the phase-space
elements evolve in classical mechanics.  
There, one may divide the classical phase-space
into small cells of different sizes at will.  
Since each cell is frozen with the 
phase-space fluid, the cell volume is conserved at all time.
By analogy, we should also let 
the spatial cell area $\theta$ 
be frozen with the incompressible flow,   
\begin{equation}
{d\theta\over dt}\equiv 
\partial_t \theta+v^i\partial_i \theta=0.
\end{equation}
This condition is exactly what we have stressed earlier
that (7) applies only in the local rest frame.

Note that the electric field vanishes
($B_{0i}'=0$) in the local rest frame of the string endpoint.  
A global reference frame that 
sees the string endpoint move at a velocity $v^j$ 
should also see an induced electric field 
$E_i=B\epsilon_{ij} v^j$.  It then follows from 
the Faraday's law,
$\partial_t {\bf B}=\nabla\times {\bf E}$, of (10) that
in this global frame
\begin{equation}
{dB\over dt}=0.
\end{equation}
It is the regime of a perfectly conducting fluid where
the "magnetic" field is frozen with the fluid.  When $\theta$
and $B$ are related by (5), eqs.(11) and (12) are consistent
to each other.  In fact, since (12) is a linear equation, this equation
also describes the evolution of $\Delta B(x,y,t)(\equiv B-B_0)$. 
At the end of this section,
we shall show that $B$ should be replaced by $\Delta B$ for the
definition of $\theta$ in (5).
  
Equation (11) provides a partial evidence 
for us to relate $\theta^{-1}\hat z$ to the hydrodynamic vorticity 
$\nabla\times{\bf v}({\bf x},t)
\equiv (\partial_1 v_2-\partial_2 v_1)\hat z$, 
which satisfies the same equation of motion.   
However, there can be infinitely many functions 
satisfying the same classical 
frozen-in equation of motion.  
Hence one needs to seek a direct
link of $\theta$ to the hydrodynamic vorticity. 
The direct link will be shown to be provided by 
a pair of canonical-conjugate fields, 
the quantization rule of which 
can be turned into the wanted link.

Proceeding along this line requires
the action of classical two-dimensional incompressible hydrodynamics 
to be identified.  We will not specify the detailed nature 
of the weak coupling among open strings, since only
the very low-energy physics of interest where the details of
interactions are smeared away.  When the interactions are 
turned on, string endpoints can move across the ${\bf B}$ field
as an incompressible fluid.
The incompressible Navior-Stokes equation reads:
\begin{equation}
\partial_t{\bf v}-{\bf v}\times\nabla\times{\bf v}=
-\nabla({{\bf v}^2\over 2}+P),
\end{equation}
where $P$ is the pressure, serving as a constraint 
to ensure the incompressibility.  Here, and from now on, all
vector products and vector differentiations refer to 
the closed-string metric $g_{ij}=\epsilon^2\delta_{ij}$
and $g_{00}=-1$.
  
Taking a divergence on (13), we see $P$ satisfy an 
instantaneous Poisson equation, so that the scalar $P$ 
is a nonlinear and non-local 
function of the pseudo-vector velocity field ${\bf v}$.  
The pressure does play a crucial role in the dynamics;
for example, though not directly appearing in the energy 
density $T^{00}={\bf v}^2/2$, the pressure contributes to
the energy flux $T^{0i}=[v^i(({\bf v}^2/2)+P)]$ 
in a crucial way.  The pressure indicated here is the dynamical
pressure and not the actual pressure of the system; the latter
is much greater than the former.  In the context of quantum
mechanics, the actual pressure is provided by the uncertainty
principle.  For example, writing the wave function of
the Schroedinger equation as $f\exp(iS/\hbar)$, the real and imaginary
parts of this complex equation are the conservation of probability
and conservation of momentum respectively.  In the momentum equation,
the uncertainty-principle pressure arises from the second derivative
of $f$, i.e., $-f\nabla^2 f/2$\cite{16}, which can be much greater than 
the kinetic energy density $f^2(\nabla S)^2/2$ when $\nabla S$ means to
describe the very slow motion.   
The non-locality and nonlinearity 
of $P$ as well as the appearance of
different parities in this fluid problem suggest 
that incompressible
hydrodynamics involves a more sophisticated 
action-principle formulation than
the potential-flow hydrodynamics, which is basically 
a problem of a single complex-scalar-field\cite{16}.  

In seeking the action of incompressible hydrodynamics, 
we begin with the local conservation laws in an ideal
fluid that contains infinitely many local invariants, 
all satisfying the frozen-in condition.  
One may choose two of these invariants
to be the new endpoint coordinate $(\alpha,\beta)$.  
This new coordinate is 
similar to the Lagrangian coordinate in fluid mechanics,
the volume element of which,
$|\nabla\alpha\times\nabla\beta|d{\bf r}^2$, 
is also a local invariant.  

The complication of this hydrodynamic
problem arises from the incompressibility condition, 
$\nabla\cdot{\bf v}=0$, which must be explicitly 
reinforced at all time. Unlike the phase-space flow,
where the incompressibility condition 
is automatically built in by its symplectic structure, 
here we need to impose a constraint 
to maintain the incompressibility.

The following Lagrangian density describes 
the evolution of the Lagrangian
coordinates $\alpha$ and $\beta$ with 
a built-in incompressibility condition:
\begin{equation}
L=-\epsilon^2[\alpha\partial_t\beta+
{(\nabla\psi+\alpha\nabla\beta)^2\over 2}],
\end{equation}
where $\alpha$, $\beta$ and $\psi$ are regarded as
independent fields and the scaling factor $\epsilon^2$
comes from $\sqrt{-Det (g_{\mu\nu})}$ in the Lagrnagian density.
This Lagrangian is obtained by
taking the non-relativistic and stiff-equation-of-state 
limits of the relativistic hydrodynamic
Lagrangian\cite{17}.  
Variations of this Lagrangian with respect to 
$\beta$, $\alpha$ and $\psi$ yield
\begin{equation}
{d\alpha\over dt}=0, \ \ \ \ {d\beta\over dt}=0, 
\ \ \ \ \nabla\cdot{\bf v}=0,
\end{equation}
where
\begin{equation}
{\bf v}\equiv\nabla\psi+\alpha\nabla\beta.  
\end{equation}
The vorticity
$\omega\hat z\equiv\nabla\times{\bf v}
=\nabla\alpha\times\nabla\beta$.
(It turns out that (14), (15) and (16) are also valid 
for three-dimensional incompressible hydrodynamics.)
As both constant-$\alpha$ and constant-$\beta$ 
lines are frozen, their intersections, 
which can be regarded as point vortices,
are also frozen with the flow.  Moreover, since
$\omega$ is the density of vortices 
and the flow is incompressible, it thus follows that
\begin{equation}
{d\omega\over dt}=
{d\over dt}(\hat z\cdot\nabla\alpha\times\nabla\beta)=0.
\end{equation}
This equation is nothing more than to say that 
the Jacobian between the effective "Lagrangian"
coordinate and the Eulerian coordinate is a local invariant.
In fact from the above
frozen-in picture, the second equality of (17) holds for
any incompressible velocity field ${\bf v}$, 
even when they are unrelated to $\alpha$ and $\beta$.
Verification of (17) requires some algebra.  
A systematic way to show (17)
is to decompose the strain tensor 
$\partial v^i/\partial x^j$ into a 
traceless component and an anti-symmetric component, 
and then make use of the 
frozen-in conditions for $\alpha$ and $\beta$ to derive (17).

We proceed to examine what the constraint field $\psi$ 
means.  Construct the energy flux $T^{0i}$ from (14)
and compare it with the $T^{0i}$ given 
four paragraphs earlier.
One finds that
\begin{equation}
{d\psi\over dt}= {{\bf v}^2\over 2}-P.
\end{equation}
Thus $\psi$ accounts, in part, 
for the accumulated effects of the
non-local pressure $P$ over the past evolution.

The Lagrangian (14) further allows us to identify 
the conjugate momentum of $\beta$,
\begin{equation}
\pi_\beta=-\epsilon^2\alpha.
\end{equation}
The fields $-\epsilon^2\alpha$ and $\beta$ are 
the canonical conjugate pair.
Upon applying the quantization rule, they satisfy
\begin{equation}
[\alpha,\beta]=i\hbar\epsilon^{-2}.
\end{equation}
Equation (20) is the desired 
condition that provides the direct link between 
$\theta$ and the vorticity.

The physical meanings of $\beta$ and $\alpha$ 
can be best illustrated 
by the stationary flows.  A two-dimensional 
stationary flow can be constructed 
using an auxiliary scalar field $\chi$, where 
${\bf v}=\hat z\times\nabla\chi$ and the 
vorticity $\omega=\nabla^2\chi$.
Since ${\bf v}\cdot\nabla\omega=0$ in a stationary 
flow, it follows that $\nabla^2\chi=U(\chi)$ 
for an arbitrary function $U$.
Thus, there exist infinitely many stationary-flow solutions, 
$\chi(x,y)$.  The simplest stationary flows 
are the planar flows and circular flows, where the velocity 
is along the direction invariant to translation and rotation, i.e., 
the (angular) velocity being a Killing vector.  In these two special 
cases, the stationary flow profile can be arbitrary.

A planar shear flow ${\bf v}=\hat y V(x)$ has 
no pressure gradient and it
yields $\beta=y-V(x) t$, $\alpha=V(x)$ and $\psi=t(V(x)^2/2)$.  
A circular shear flow
${\bf v}=\hat\phi r\Omega(r)$, on the other hand, 
gives $\beta=\phi-\Omega(r)t$, $\alpha=r^2\Omega(r)$ and 
$\psi=t[(r^2\Omega^2/2)-\int r\Omega^2 dr]$.
It is clear that $\beta$ is nothing but the co-moving coordinate 
in the direction of the flow;
the flow speed ($\alpha$) is the negative conjugate momentum.  
We shall, from now on, call the canonical coordinate ($\alpha,\beta$) to
be the Lagrangian coordinate for convenience, though the Lagrangian
coordinate in fluid mechanics has a slightly different meaning.

We are now ready to relate 
$\omega$ to $\theta^{-1}$.
Expressing $\alpha=\alpha(x)$, $\beta=y-t\alpha(x)$ 
for the planar shear flow, we find from (7) and (20) that
\begin{equation}
[\alpha,\beta]={d\alpha\over dx}[x,y]=i\omega\theta
=i\hbar\epsilon^{-2},
\end{equation}
where the commutator is defined in terms of 
the star products (8). The first equality of (21) is
valid only up to $O(\theta^2)$.  Note from the third
equality of (21), if both $\omega$ and $\theta$ assume
classical values, the small parameter $\epsilon$ will be
of quantum origin, as it would have scaled as $\sqrt{\hbar}$.

One may, in fact, do better than the $O(\theta^2)$ 
order for (21), where
all high-order terms vanish. 
This can be shown by changing the Cartesian coordinate 
${\bf x}=(x,y)$ to the Lagrangian coordinate 
${\bf\eta}=(\alpha,\beta)$.  With the latter 
coordinate, the new noncommutative parameter 
becomes $\hbar\epsilon^{-2}$.  The star product for
the Lagrangian coordinate assumes 
the same conventional form:
\begin{equation}
f({\bf\eta})*h({\bf\eta})
=\exp[i(\hbar/2\epsilon^2)\epsilon^{ij}\partial_i^a\partial_j^b]
f({\bf\eta}_a)h({\bf\eta}_b)|_{{\bf\eta}_a={\bf\eta}_b={\bf\eta}}.
\end{equation}
We now have $x=x(\alpha)$ and $y=\beta+t\alpha$.
A straightforward calculation shows that
$[x,y]$ is terminated beyond $O(\hbar)$ in 
the series expansion of the star product
(22), and $[x,y]=i\hbar\epsilon^{-2}\omega^{-1}$.  
By definition, we also have $[x,y]=i\theta$.  
Hence (21) is valid for all orders of $\theta$, 
and the vorticity $\omega$ can be identified 
to be $\hbar\theta^{-1}\epsilon^{-2}$ in a planar flow.  

A similar result also holds for the circular shear flow.
Originally,
\begin{equation}
[\alpha,\beta]=i\theta\omega=i\hbar\epsilon^{-2}
\end{equation}
up to $O(\theta^2)$.
One now changes the polar coordinate 
$(r^2/2,\phi)$ to again the Lagrangian coordinate 
${\bf\eta}=(\alpha,\beta)$, 
where the coordinate transformation
reads $r^2=r^2(\alpha)$ 
and $\phi=\beta+t[\alpha/r^2(\alpha)]$.
It follows that 
$[r^2/2,\phi]=i\hbar(rdr/d\alpha)=i\hbar\epsilon^{-2}\omega^{-1}$, 
valid for all orders of $\hbar$.  By definition 
$[r^2/2,\phi]=i\theta$.  
Hence (23) is also valid for all orders
of $\theta$, and again $\omega=\hbar\theta^{-1}\epsilon^{-2}$ 
in a circular flow.

One important conclusion derived from this section is that
$\omega$ is proportional to $\theta^{-1}$.  In general, the $\omega$ field
can be rather localized in space and does not need to have a strong
uniform background, which corresponds to a rigid-body rotation.  It
thus follows that the assumed strong uniform compoenent of the 
NS field, $B_0$, has disappeared in this problem.  The relevant quantity 
described by (12) is instead the non-uniform component $\Delta B$.
It is $\Delta B$ that is proportional to a localized $\omega$ and
replaces $B$ in (5) for the definition of $\theta$.

In the above, we have used specific examples to identify $\omega$
with $\theta^{-1}$.  However, the fact that
$\omega=\hbar\theta^{-1}\epsilon^{-2}$ to all orders holds 
only for a very specific
coordinate for a given $\theta(x,y)$.  In general this relation
is valid only at the Poisson's level,
i.e., to the lowest-order star-product expansion.
Due to the existence of a $\theta$-gradient, it makes
the commutator between the two space components of one set of
coordinate, e.g, $[x,y]$, different
from that of another set, e.g., $[\bar x,\bar y]$,
by an amount $O(\hbar^3)$.  
These high-order quantum effects break the
area-preservation condition for different sets of coordinates, which
are classically related by area-preserving transformations.
This feature is new and generated by the $\theta$-gradient.  It is 
explored below.

\section{Quantum Area-Preserving Maps and Solitons}

Area-preserving maps connect an equivalent class of 
coordinates, in the sense that a function expressed in terms of
one set of coordinate possesses the same properties as that
in terms of another set of coordinate of the equivalent class. 
In the commutative space, the area-preserving 
map is a transformation that maps every local
point in the old frame to a new frame in an area
preserving manner.  However, in the noncommutative 
space, there no longer exist local points but unit cells of
finite areas, as a result of the space quantum fluctuations.  
Thus, the quantum area preservation 
involves non-local conditions when measured in the classical space.
When all unit cells have the same
areas, i.e., a constant $\theta$, the quantum
area-preserving maps can  
still retain those good properties of the classical area-preserving
maps.  To be specific, the coordinates are operators in the 
noncommutative space and a well-defined function 
of space coordinates in the commutative space can
become ill-defined in the noncommutative space, because
a function of operator coordinate depends also on the 
ordering of operators, a feature that
reflects the fuzziness of the noncommutative space.
However, once a function is well-defined with one set of operator 
coordinate, it will remain so with another
set of operator coordinate, provided that these
coordinates are connected by an area-preserving 
transformation.  The canonical coordinates in
conventional quantum mechanics provide an example of such.
However, when the unit
cells have various sizes, their quantization can
be a subtle problem\cite{18}.  It turns out that the
Lagrangian coordinate can help solve this problem.

The examples given in the last section 
have demonstrated the usefulness of 
the Lagrangian coordinates,
which turn complicated star-product calculations 
with a non-uniform $\theta$
into the simpler one with a constant noncommutative parameter
$\hbar\epsilon^{-2}$.
The Lagrangian coordinates also
allow for construction of quantum area-preserving, 
or canonical, transformations in a systematic way.
A canonical transformation transforms 
the canonical coordinate $(\alpha,\beta)$ to
a new canonical coordinate $(\bar\alpha,\bar\beta)$, such that
$[\bar\alpha,\bar\beta]$ remains to be $i\hbar\epsilon^{-2}$.
Much like in the commutative space, a finite quantum canonical 
transformation can be generated by successive 
applications of many infinitesimal canonical transformations.
Infinitesimal transformations of the form, 
$\bar\alpha=\alpha-\partial\delta S/\partial\beta$ and
$\bar\beta=\beta+\partial\delta S/\partial\alpha$,
are canonical, where $\delta S$ is the generation function,
an analytical function of $\alpha$ and $\beta$. 
One can easily show that 
$[\alpha,\delta\beta]+ 
[\delta\alpha,\beta]=0$, so that they 
indeed are canonical.  

Though being useful for its 
the symplectic structures, the Lagrangian coordinate is 
a deformed coordinate after all. 
Moreover, the Lagrangian coordinate describes the
non-inertial frame that
accelerates/decelerates with the fluid elements,
thereby yielding a time-dependent 
metric $g_{ij}'(\alpha,\beta,t)$ and 
$g_{i0}'(\alpha,\beta,t)$, even in the presence of
a stationary flow.  Therefore, to understand
the space quantum fluctuations, it is more relevant to
examine the quantum fluctuations of
the inertial-frame coordinate, described 
by the static metric $g_{ij}=\epsilon^2\delta_{ij}$,
than those of the Lagrangian coordinate.

In the noncommutative space, the static
coordinates, such as the Cartesian coordinate, can be canonical
coordinates when $\theta^{ij}$ is uniform, and hence 
different canonical coordinates can be generated by
area-preserving maps.  A function expressed in terms 
of these different canonical coordinates is equally
well-defined.  However,
when $\theta^{ij}$ is non-uniform,
all inertial-frame coordinates  
become non-canonical, i.e., the commutator not being a constant.  
As a consequence, quantum area-preserving 
maps among these inertial-frame coordinates 
generally do not exist, resulting in that functions
expressed in terms of the inertial-frame coordinates
become coordinate-dependent.  

This peculiar feature implies
that different inertial-frame coordinates suffer from quantum
fluctuations in different ways, and hence 
they become not equivalent to each other.  From this 
observation, it now becomes clear that area preservation
imposes a subtle condition to ensure the quantum
fluctuations of coordinates to be transformed in the same manner
as the classical coordinates.  
That is, the Lagrangian coordinate can package
all high-order quantum effects of
star product in a neat manner, as if the quantum fluctuations
were non-existent.

Despite such an unpleasant feature of the 
inertial-frame coordinates, 
it turns out that there exists a class of non-uniform 
$\theta^{ij}$, or vortex flows, that can make the high-order
quantum fluctuations in the inertial-frame coordinates
vanish, and therefore area-perserving maps can be restored.  
These special flows are classical objects that
survive in the noncommutative space.  
Such vortex flows will be called the
"soliton" solutions and shown to be the solutions
of the Dirac-Born-Infeld Lagrangian\cite{18}.  Below, we
explore them.

Among all possible choices of inertial-frame coordinates, 
the holonomic and orthogonal coordinates are 
suited for describing the global space support of 
a function, and hence can naturally be extended to the
operator coordinates in the noncommutative space.  
The most natural holonomic and orthogonal
inertial-frame coordinate is the Cartesian coordinate.  
In the two-dimensional {\it commutative} space, 
the Cartesian coordinate
can be transformed to one and only one holonomic  
and orthogonal coordinate in an area-preserving manner;
it is the special polar coordinate $((r^2+r_0^2)/2,\phi+\phi_0)$,
where $r_0^2$ and $\phi_0$ are constants.
One may prove this statement by showing: (a)
one component, e.g., $f$, of the transformed coordinate must have 
a dimension, (length)$^2$, and the other component, e.g., $h$,
a dimension, (length)$^0$; (b) with 
$f=r^2\tilde f(\phi)$ and $h=h(\phi)$, the desired
result follows.  

We shall therefore be confined to examining under what conditions 
the quantum area-preserving transformation between the special
polar and Cartesian coordinates
holds in the noncommutative space.
A class of stationary flows of radial profile,
$\omega=\omega(r)$, 
will be examined in details.  
We have the commutator of the special polar coordinate,
$[(r^2+r_0^2)/2,\phi+\phi_0]=[r^2/2,\phi]$, 
and it has been shown earlier that
$[r^2/2,\phi]=i\hbar\epsilon^{-2}\omega^{-1}(r)$ 
to all orders of $\hbar$.  Let $\theta$ be defined by
(7), we next explore how to make $\theta$ equal to
$\hbar\epsilon^{-2}\omega^{-1}$.

A convenient basis for
evaluating $[x,y]$ is again the Lagrangian coordinate, 
${\bf\eta}=(\alpha,\beta)$, where $x$ and $y$ are written as
\begin{equation}
x({\bf\eta}_a)=r(\alpha_a)
\cos[\beta_a+{t\alpha_a\over r^2(\alpha_a)}], \ \ \
y({\bf\eta}_b)=r(\alpha_b)
\sin[\beta_b+{t\alpha_b\over r^2(\alpha_b)}].
\end{equation}
Note that $x_a y_b=r(\alpha_a)r(\alpha_b)\cos(q_a)\sin(q_b)
\propto(1/2)(\sin(q_+)+\sin(q_-))$,
where $q_+\equiv q_a+q_b$ and $q_-\equiv q_a-q_b$, with
$q\equiv\beta+t\alpha/r^2(\alpha)$.  Hence
$x_a y_b$ contains four terms proportional to 
$\exp(\pm iq_+)$ and $\exp(\pm iq_-)$. 
Moreover, the operator
$\exp[i(\hbar\epsilon^{-2}/2)\epsilon^{ij})\partial_i^a\partial_j^b]$ 
of the star product in (22) can be re-expressed as 
$\exp[i\hbar\epsilon^{-2}(\partial_{\alpha_-}\partial_{\beta_+}-
\partial_{\alpha_+}\partial_{\beta_-})]$.  
When $x$ and $y$ are
substituted for $f$ and $h$ in (22), 
one may perform Fourier 
transformations for $\beta_+$ and $\beta_-$ 
and generates four Fourier modes for
the ordinary product $x_a y_b$.  

The star-product for these modes has the form
\begin{equation}
e^{\pm[i\beta_-+\hbar\epsilon^{-2}(\partial/\partial\alpha_+)]}
F(\alpha_a,\alpha_b)+
e^{\pm[i\beta_+-\hbar\epsilon^{-2}(\partial/\partial\alpha_-)]}
H(\alpha_a,\alpha_b).
\end{equation}
The operator, 
$\exp[\pm\hbar\epsilon^{-2}(\partial/\partial\alpha_+)]$, 
is simply a shift operator that shifts $\alpha_+$ in $F$ 
by $\pm\hbar\epsilon^{-2}$.
Likewise, $\exp[\mp\hbar\epsilon^{-2}(\partial/\partial\alpha_-)]$ 
shifts $\alpha_-$ in $H$ by $\mp\hbar\epsilon^{-2}$.
It thus follows 
\begin{eqnarray}
x*y=&&{i\over 4}[r^2(\alpha+{\hbar\over 2\epsilon^2})-
r^2(\alpha-{\hbar\over 2\epsilon^2})]\nonumber\\
&&-{1\over 2}r(\alpha+{\hbar\over 2\epsilon^2})
r(\alpha-{\hbar\over 2\epsilon^2})
\sin[2\beta+t({2\alpha+\hbar/\epsilon^2\over
2r^2(\alpha+(\hbar/2\epsilon^2))}+
{2\alpha-\hbar/\epsilon^2\over 2r^2(\alpha-(\hbar/2\epsilon^2))})],
\end{eqnarray}
and hence
\begin{equation}
[x,y]={i\over 2}[r^2(\alpha+{\hbar\over 2\epsilon^2})
-r^2(\alpha-{\hbar\over 2\epsilon^2})].
\end{equation}

Apparently, $[x,y]\neq [r^2/2,\phi]$.  If we define $[x,y]$ to be
$i2\pi B^{-1}$, c.f., (7), then the background field $B$ is 
not related to the vorticity
$\omega(=2(d r^2/d\alpha)^{-1})$ by $\hbar B/2\pi\epsilon^2$, 
but by the non-local condition given on the right-hand side of (27).  
At the Poisson level, i.e., 
to the leading order in $\hbar$, we indeed have 
$i\theta=[x,y]\approx (i\hbar/2\epsilon^2)
dr^2/d\alpha=i\hbar\epsilon^{-2}\omega^{-1}=[r^2/2,\phi]$,
and hence $\omega\approx \hbar/\theta\epsilon^2$.  
The deviation from canonical
condition between the polar and Cartesian coordinates, or from
$\omega=\hbar/\theta\epsilon^2$, starts from $O(\hbar^3)$
and only odd powers of $\hbar$ contribute to the difference.  
The deviation is of quantum origin since it involves $\hbar$.  

The soliton solutions are those classical vortex flows that
can survive in the noncommutative space by making $[x,y]=[r^2/2,\phi]$,
or equivalently making $\omega=\hbar/\theta\epsilon^2$. 
These soliton solutions satisfy 
$[x,y]=(i\hbar/2\epsilon^2)d r^2/d\alpha$, thus
obeying a linear differential-difference equation:
\begin{equation}
\hbar\epsilon^{-2}{d r^2(\alpha)\over d\alpha}=
r^2(\alpha+{\hbar\over 2\epsilon^2})-
r^2(\alpha-{\hbar\over 2\epsilon^2}).
\end{equation}
The solution to this equation is not unique.  Due
to the linearity of (28), all solutions can be
superposed.  One exact solution 
that can be approximated by the increasingly 
higher-order expansion of 
small-$\hbar$ satisfies $d^n r^2/d\alpha^n=0$ for
$n\geq 3$.  This solution terminates the 
$\hbar$-expansion and
only the $O(\hbar)$ term survives  
It yields that $r^2=c_2\alpha^2+c_1\alpha+c_0$, or
\begin{equation}
\alpha=a\sqrt{1\pm (r/r_0)^2}+b,
\end{equation}
where $r_0$, $a$, $b$ and $c$'s are all constants.
Though the Planck constant $\hbar$ appears in (28),
it disappears in the solution (29), indicative of
these soliton solutions to be of classical origin.

Note that $\alpha$ is the angular momentum, 
and a finite-angular momentum
near $r=0$ produces a singular flow 
$V_\phi\sim r^{-1}$.  To avoid the singular
behaviors, we may choose the plus sign 
in the square-root of (29) and 
let $b=-a$, thus obtaining
\begin{equation}
\alpha(r)=a(\sqrt{1+({r\over r_0})^2}-1), \ \ \ 
\omega(r)={a\over r_0\sqrt{r_0^2+r^2}}, \ \ \
G_{ij}(r)=(\alpha'\epsilon)^2
{a^2\over 4\pi^2\hbar^2 r_0^2(r_0^2+r^2)},
\end{equation}
where Eq.(1) in the decoupling limit has 
been used to obtain the open-string metric 
$G_{ij}$, given that $g_{ij}=\epsilon^2\delta_{ij}$.
Note that if all flow quantities and the B-field are of
classical values, we need $\epsilon^2$ to scale as
$\hbar$, thereby yielding a finite classical $G_{ij}$ 
of order $\hbar^0$ and $\epsilon^0$.

It turns out that the other choice of sign in 
the square-root of (29)
can also make sense.  Here we may alternatively have
\begin{equation}
\alpha(r)=a(\sqrt{1-({r\over r_0})^2}-1), \ \ \ 
\omega(r)={-a\over r_0\sqrt{r_0^2-r^2}}, \ \ \
G_{ij}(r)=(\alpha'\epsilon)^2
{a^2\over 4\pi^2\hbar^2 r_0^2(r_0^2-r^2)}.
\end{equation}
This is a confined flow, where 
the vorticity diverges at the 
rotating boundary $r=r_0$.  

Equation (30) shows a vortex solution that 
has a rigid-body rotation ($V_\phi\sim r$)
at $r\ll r_0$ and exhibits a flat rotation 
($V_\phi\to const.$) at $r\gg r_0$. 
Interestingly, in the limit $r_0\to 0$, 
we have the open string metric
$G_{ij}\propto r^{-2}$, making $G_{ij} dx^i dx^j
=(d\ln r)^2+d\phi^2$.  Since
$\phi$ is periodic and $-\infty<\ln r<\infty$, 
the topology of the open string metric becomes
a cylinder.  In other words, the open-string 
and closed-string metrics are both 
flat in this limit, but with different 
topologies as a result of 
the presence of a soliton vortex.

On the other hand, the open-string metric 
satisfies $G_{ij}\propto (r_0^2-r^2)^{-1}$ 
for the other $\alpha$
given by (31).  This soliton solution 
can have an interesting connection to magnetic
monopoles.
The open-string spatial line element 
can be expressed as:
$dl^2=r_0^2[d\xi^2+\tan^2(\xi) d\phi^2]$, where 
$r\equiv r_0\sin\xi$, and 
the non-relativistic kinetic energy is nothing but 
\begin{equation}
T={r_0^2\over 2}[\dot\xi^2+\tan^2(\xi)\dot\phi^2]=
{r_0^2\over 2}[\dot\xi^2+\cot^2(\xi)p_\phi^2],
\end{equation}
where $p_\phi$ is the $\phi$-momentum.  
This form of kinetic
energy may be derived from the metric 
of a symmetric rotor in three dimensions:
$dl^2=d\xi^2+\sin^2(\xi)d\phi^2+
s^2(d\psi-\cos(\xi)d\phi)^2$,
where $s$ is a real constant.  Again, express the 
non-relativistic kinetic energy from this line element.  
Upon recognizing the momenta 
$p_\psi=s^2(\dot\psi-\cos(\xi)\dot\phi)$
and $p_\phi=(\sin^2(\xi)+s^2\cos^2(\xi))\dot\phi
-s^2\cos(\xi)\dot\psi$, we find that
\begin{equation}
T={1\over 2}[\dot\xi^2+
{(p_\phi+\cos(\xi)p_\psi)^2\over\sin^2(\xi)}
+s^{-2}p_\psi^2].
\end{equation}
In the case $p_\psi\gg p_\phi$ or $p_\phi=0$, 
(33) is reduced to (32) apart from a zero-level energy
$s^{-2}p_\psi^2$, which can be made arbitrarily small
when $s^2\to\infty$.  Note that the $p_\psi\gg p_\phi$ 
limit is reminiscent
of the decoupling limit taken for the string metric,
where the particle gyro-motion is negligible 
and the actual 
motion is governed by the diamagnetic current.
Therefore, the dynamics in the 
open-string metric constructed by the
vortex solution (31) is equivalent to the 
low orbital-angular-momentum dynamics of 
a three-dimensional rotor, for which 
the "spin"($\psi$)-degree of freedom provides 
an effective background field.  In fact, 
if we ignore the last term of (31) by taking 
$s^2\to\infty$, the dynamics is 
identical to the one around a monopole
of magnetic charge $Q_m(\propto p_\psi)$.
 
These solutions given in (30) and (31) are the classical ones,
as all high-order terms of the star product vanish.
Somewhat surprisingly, the vorticity $\omega$ found here turns out
to be the electric-like classical gauge-field solutions,
$F_{0i}$, of
the Dirac-Born-Infeld (DBI) Lagrangian\cite{9,19}, 
$L\sim\sqrt{-Det(g_{\mu\nu}+\alpha' F_{\mu\nu})}$,
created by a delta-function 
point source in the commutative space.  

One may derive the 
Dirac-Born-Infeld solution straightforwardly
by expressing $F_{0i}=-\partial_i\xi(r)$, where
$d\xi/dr$ is to be identified as $\omega$.  Variation with
respect to $\xi$ yields the equation of motion:
\begin{equation}
{d\over dr}[{(\alpha')^2
r\xi'\over\sqrt{\epsilon^2-(\alpha'\xi')^2}}]=0, 
\end{equation}
where
$\xi'\equiv d\xi/dr$.  Taking the squared bracket to be a
finite constant is equivalent to having a delta-function source
at $r=0$. Solutions (30) and (31) can be obtained from
this procedure, and they correspond to having 
real and imaginary sources, respectively.  
In this decoupling limit, where $\alpha'\to \epsilon\to 0$,
the DBI gauge-field solution typically has a finite amplitude 
on the order of $(\alpha')^0$, and a length scale of order unity
when the source strength also scales as $\epsilon$.
The $(\alpha')^0$ amplitude scaling is indeed what has been assumed
for $B$ or $\theta$ in the decoupling limit.
Thus, the DBI solutions recover the correct scaling 
and are indeed valid classical solitons. We shall briefly 
comment on this point below.

These vortices studied here are the NS fields which obey the
equation of motion (17) or $\omega=\nabla^2\chi=U(\chi)$ of
an arbitrary function $U$ in steady states, not compatible with
the DBI gauge field that obeys a different equation of motion derived
from the DBI Lagrangian.  
Only for some special choices of $U(\chi)$, 
the two give the same solutions,
which survive the quantum fluctuations. In the present case, we have a 
rather complicated $U(\chi)$, for which the inverse function of $U(\chi)$
is $\chi=U^{-1}(\omega)=\omega^{-1}-(1/2)\ln(1\pm\omega^{-1})$,
where the $"+"$ and $"-"$ signs refer to solutions (30) and (31),
respectively. 

Another example that exhibits this feature
is a magnetic-like DBI field, for which we need to represent
$B_{12}(r)=d(r A_{\phi})/d(r^2)$ in the DBI Lagrangian.  
Redefine $\bar\xi=r A_{\phi}(r)$,
and perform the variation with respect to $\bar\xi$.  With a point
source, we find that $B_{12}$ is a constant.  Again when the source
strength is of order $\epsilon$, we recover a $\theta$ of correct
$\epsilon$ scaling.  This result recovers the familiar 
constant-$\theta$ case,
where the high-order quantum fluctuations of the flat-space 
coordinates vanish.  This constant $\omega$ solution corresponds 
to choosing a trivial $U(\chi)=const.$.

Indeed, it has been known that some classical DBI solutions can be
ones that survive in the noncommutative space. 
They make the string world sheet conformal, whereby quantum corrections 
of high order vanish.  The soliton solutions found here are also 
classical DBI solutions, and they make the two dimensional space in the  
D-brane world volume area preserving, thereby nullifying the
space high-order 
quantum corrections.  The connection between the two is not
obvious since the string
world sheet is outside the D-brane and involves two string endpoints,
but the D-brane 2-sheet is within the brane.  In addition, the area
preservation and the space conformality are exclusive properties of
coordinate transformations.
Thus the two seem to be orthogonal to each other, and yet they give
rise to similar results.  There may be duality-like connections between
those DBI solutions and the vortex solitons constructed by our way, but
the connections are unclear at the moment.

\section{2D Dynamics of a Strongly Magnetized 3D Electron Gas}

One of the known classical three-dimensional systems 
that exhibit two-dimensional incompressible
hydrodynamic behaviors is the strongly magnetized 
three-dimensional 
electron gas\cite{15}.  The system is moderately long in the direction
of magnetic field, with the electrons distributed uniformly along
the magnetic field.  The gyro-radius of electron
is so small that only the guiding-center motion 
is relevant for the large-scale,
low-frequency motion across the magnetic field.  
When the electron density is high, 
even a small amount of charge 
inhomogeneity is sufficient to produce a
sizable electric field, which gives rise to electron drift motion 
perpendicular to the magnetic field with a velocity:
\begin{equation}
{\bf v}=c{{\bf E}\times {\bf B}_0\over B_0^2},
\end{equation}
where ${\bf E}$ is the electrostatic electric field
$-\nabla\Phi$,
and ${\bf B}_0$ is the strong and uniform background 
magnetic field.  

Take a divergence over (35), it shows that the velocity satisfies the
incompressibility
condition, $\nabla\cdot{\bf v}=0$.  On the other hand, the vorticity 
\begin{equation}
\omega=\hat z\cdot\nabla\times{\bf v}=
c{\nabla^2\Phi\over B_0},
\end{equation}
and is proportional to the electron 
density excess/depletion $\delta\rho_e$.  
The vorticity $\omega$ has the same sign as
the gyro-rotation for electron density excess,
but opposite for density depletion (electron holes).

On the other hand, with incompressibility, 
the electron density excess/depletion $\delta\rho_e$ obeys 
the frozen-in condition.  Therefore, from (36)
the vorticity $\omega$ also obeys the frozen-in
condition.  It means that the collective 
equation of motion for these strongly magnetized
electrons is the vorticity equation (17),
or equivalently the Navior-Stokes equation.
The example illustrates that in this coupled 
many-body system, the low-energy effective theory 
is the Navior-Stokes theory 
and not the original Newton-Maxwell theory.
The electron system behaves like a neutral fluid is also
evidenced by a comparison of the electric and kinetic
energy densities.  The former is $E^2/8\pi$ and the latter
$n_e m_e v^2/2=(E^2/8\pi)(4\pi n_e m_e c^2/B^2)\gg E^2/8\pi$
for a dense electron system in a very strong magnetic field, so
strong that the magnetic energy density $B^2/8\pi$ is much larger
than above two.  
Notice that no collisions, but small long-ranged electric forces, are
needed to make electrons behave like an incompressible fluid.
This is because
the strong magnetic field has provided large pressure
to yield a very stiff electron equation of state.    

The electron gas exhibits quantum effects only
at low temperature, and a cooled semiconductor in a strong magnetic
field may be made to produce the quantum vortices studied here. 
This electron-gas
system is similar to the one that shows quantum Hall effects, except that
the latter has a small dimension along the magnetic field, i.e., 
monolayer systems, whereas the present one needs a finite longitudinal
dimension.  Although both systems exhibit two-dimensional dynamics,
there is an important difference.  The transverse electric-field, 
which is essential to produce vortex motion, is long-ranged for a charge
colomn, but becomes short-ranged for a charge layer.  The difference
is reflected by the fact that the electric fields are mostly
perpendicular both to the charge column and to the charge layer; 
the former
lies perpendicular to the magnetic field but the latter aligns with the
magnetic field, and therefore the spatial dependence of 
their transverse electric
fields is different.

To make a comparison with what has been presented above, 
we note that given a
strong background $B_0$ field, the change of 
magnetic field $\delta B$ is negligible, so as not to
over-pressure the electron gas for the slow vortex motion.  
Typically $\delta B/B_0$ is about the ratio
of the electron gyro-period to the vortex rotation period, 
and hence in this strong-field regime, 
it is the vorticity, or the electron density perturbation
$\delta\rho_e$ (c.f.,(36)), that plays the role of 
$\theta^{-1}$ in our previous discussions.  

As discussed in Secs.(3) and (4), the Lagrangian coordinate
$(\alpha,\beta)$ is useful for incorporating all quantum effects
in the many-body system through the star product. Elemental vortices
can be excited in this electron-gas column, and they are quantum 
objects different from the classical solitons given in (30) and (31).
The elemental excitations tend to have angular momenta on the
order of $\hbar$ but the classical solitons have an angular momentum
of many $\hbar$.  The fact that static inertial-frame 
coordinates are generally
non-canonical, due to the high-order quantum corrections, 
may produce interesting experimentally detectable features.
However in the following
discussions, we shall make no attempt to examine the detailed 
quantum effects, but simply to predict a leading-order 
feature of these quantum vortices.  Much like in the classical
system, the quantum vortex is also charged; its charge can be quantized,
but in a way different from that arising from the quantum Hall effect.

From (36), it follows that quantization
of angular momentom, $L=m_e\int d^2{\bf r}\omega$, 
can be made equivalent to quantization 
of electric charge per unit length, 
$Q_e/l_\parallel=\int d^2{\bf r}\nabla\cdot{\bf E}/4\pi$, along the 
background field ${\bf B}_0$, and they are related by 
\begin{equation}
Q_e/e={\pm 137 n\over 2}({l_\parallel\nu_{cy}\over c}),
\end{equation}
where $l_\parallel$ is the system dimension 
along the magnetic field,
$\nu_{cy}=eB_0/2\pi m_e c$ the electron cyclotron frequency with $m_e$
being the electron mass, and
the positive integer $n\equiv \pm L/\hbar$.  

A estimate of relevant parameters
can be in order.  For a magnetic field $B_0= 10^5$Gauss and
the system size $l_\parallel= 10^{-3}$cm,
we have $Q_e/e\approx\pm 1.37n/2$.  Since the amount of charge
is contributed by all electrons along the vortex coloumn of size
$l_\parallel$, this length must be smaller than or comparable to 
the electron coherent length.  At a sub-Kelvin temperature,
the electron coherent length can be as large as $10^{-3}$cm in
a quality semiconductor sample, and such a quantum-vortex system can be
made to exist.

For elemental vortex excitations, $n$ is on the
order of unity, the field strength $B_0$ and longitudinal system size
$l_\parallel$ given above yield that the vortex quasi-particle 
contains a charge $Q_e\sim e$, which can even be a fraction of $e$.
Notice that the elemental vortex excitations are expected to be
extended objects, covering a transverse size
much greater than both electron mean separation and electron
gyro-radius $r_e(\equiv\sqrt{\hbar c/eB_0})$ 
at the first Laudau level.  For the vortex
dynamics to be close to two-dimensional, it also needs $l_\parallel$ 
still much greater than the transverse vortex size.  
With the field strength $B_0=10^5$Gauss, we have the electron
gyro-radius $r_e\sim 10^{-6}$cm at the first Landau level, which
is indeed much less than the longitudinal system size 
$l_\parallel$.  The vortex transverse size thus lies
in between $10^{-6}$ and $10^{-3}$cm on the sub-micron 
to micron scales.

With this prediction, we are ready to compare this
column electron-gas system with the layer electron-gas system
that shows quantum Hall effects.  Due to the short-range nature
of the transverse electric field, the layer electron gas  
does not behave like
a neutral fluid as the column electron gas does.  Susskind 
investigated quantum Hall effects in connection with the noncommutative
space by formulating the compressional displacement as the relevant gauge
field\cite{20}, in contrast to the vorticity proposed in this work.  The 
elemental excitation in the quantum Hall system is the charge hole,
which is created, in Susskind's formulation, by strongly rarefying
the local electron density, a mechanism to be contrasted with the
present case where the electron density perturbation in a vortex 
is of small amplitude and extends over a larger area.
Over one angular momentum quanta, Susskind's charge hole contains
an integer fraction of electron charge, whereas in the present 
case we have an electric charge quantized directly not to the electron
charge and having a dependence on the system length along the magnetic
field.

\section{Discussions and Conclusions}

It is worthwhile to reiterate that incompressible 
hydrodynamics in the noncommutative space 
is a low-energy effective theory, where the
the background Neveu-Schwarz $B$ field is replaced by the 
vorticity field $\omega$ and thus
$\theta\approx\hbar/\omega\epsilon^2$.  To put it
in the context of the original NS field, it is the inhomogeneous
component of $B$, i.e., $\Delta B(=B-B_0)$, that replaces $B$ in
the definition of $\theta$ in (5).  As $B_0\gg\Delta B$,
the non-commutative parameter $\theta$ becomes much bigger than
was originally defined in (5).  This modification of 
the NS field should be due to the collective
effect of $N$ interacting open strings that nullifies the original
strong homogeneous $B_0$ background.  The case study for the 
semicondutor given in Sec.(IV) provides an evidence for such an 
assessment.

The appearance of vortices
can be analogous to the appearance of a new gauge field
in an effective field theory.
In fact, other than vortices classical hydrodynamics 
also supports pressure waves, i.e., the sound waves, 
that exist in the linear regime and 
are of high frequency.  
These high-frequency degrees of freedom are supported 
by the thermal pressure (or magnetic pressure for the case
considered in Sec.(IV)), which is much greater than
the nonlinear dynamical pressure $P$ addressed in Sec.(III).
These degrees of freedom
have been filtered out (or integrated out in the semantics of
field theory) in incompressible hydrodynamics, 
which deals only with low-frequency vortices.
In many respects, the classical
incompressible hydrodynamics indeed resembles the effective 
gauge-field theory of a coupled many-body system.
High-energy excitations in quantum-field theories are 
the counterpart of sound waves, and
upon integrating out high-energy contributions, 
the low-energy effective field is a new gauge field.
Such an effective gauge field corresponds to the 
low-frequency vortices studied here.  

In connection with the above general picture,
we have also examined the strongly magnetized electron
gas column in Sec.(IV).  This electron system indeed exhibits 
two-dimensional incompressible vortex motion,
due partly to the presence of a strong background 
magnetic field, which provides strong pressure 
to produce a stiff equation of state, 
and partly to the long-ranged, mean-field 
Coulomb interactions, which give rise 
to the $E\times B$ drift motion.

Though we have addressed various aspects of vortices, 
in the context of the string endpoint dynamics in a
D3-brane, the much more delicate issue as to how these
vortices are excited has not been touched upon. 
This aspect of problem 
can nevertheless be qualitatively understood
from the classical physics as well.  
Vortices are intrinsically nonlinear
objects, absent in the linear regime 
of hydrodynamics when there 
exists no background rotational flow.  
They can initially be excited
through nonlinear sound interactions.
However, sound waves are scalar fields 
but vortices consist of pseudo-vector
fields.  The excitation of the latter 
requires breaking of local chiral 
symmetry but with the global chiral symmetry 
to remain intact.  Hence, vortices
must be excited at least in pairs of 
opposite chirality.  When vortices become
densely populated in the fluid system, 
annihilation and formation of vortices
through vortex-vortex interactions 
become the dominant processes.  Moreover,
vortices tend to merge to form larger ones.  
In a relaxed system, all small
vortices tend to merge into a single big vortex, 
a manifestation of the Bose-Einstein condensation.  
Though the above description portrays the
vortices of classical hydrodynamics, 
there is no reason to believe the quantum 
vortices should qualitatively behave differently. 
In particular, the soliton solutions given in (30)
and (31) can be the Bose-Einstein condensates
in relaxed systems.

In sum, we have shown in this paper that the 
frozen-in, or Lagrangian, coordinate 
forms a pair of canonical conjugate fields, which can describe
the open-string endpoint dynamics via two dimensional 
incompressible hydrodynamics.  The vorticity $\omega$ of
hydrodynamics is shown to replace the background NS-field to the lowest
order of the noncommutative parameter $\theta$ when $\theta$
is non-uniform and time-dependent. Despite the 
complications arising from the gradient of the 
noncommutative parameter, we
have shown that in the Lagrangian coordinate, $\epsilon^2\theta$
is replaced by a new noncommutative parameter $\hbar$, 
the Planck constant, thereby restoring the noncommutative space to the
constant-$\theta$ geometry.

We have also identified the classical vortex flows that survive in the
noncommutative space, and they turn out to be the classical solutions to
the Dirac-Born-Infeld Lagrangian living only on the D-brane.
As some Dirac-Bohr-Infeld gauge field can indeed
have a similar property in making all quantum corrections vanish,
we suspect that the two have deep connections, a good understanding of which 
may shed lights on gauge-field theories in noncommutative geometry 
with a non-uniform $\theta$.

Finally we have also made a prediction for the existence of, 
and the quantized charge contained in, the quantum column
vortices.  These quantum vortices
can be present in a strongly magnetized electron gas in 
a semiconductor of finite thickness at a sub-Kelvin temperature.
 
I would like to thank P.M. Ho for helpful and insightful
discussions during the course of this work.  The support from
the National Science Council of Taiwan under the grant,
NSC-90-2112-M-002, is acknowledged.


\begin{thebibliography}{99}
%
\bibitem{1} E. Witten, Nucl. Phys. {\bf B 268}, 253, 1986

\bibitem{2} A. Connes, M.R. Douglas and A. Schwarz, 
Report No. JHEP 9802:003, 1998;
M.R. Douglas and C. Hull, Report No. JHEP 9802:008, 1998,
hep-th/9711165 


\bibitem{3} C. S. Chu and P.M. Ho, Nucl. Phys. {\bf B 550}, 151, 1998;
C.S. Chu and P.M. Ho, Nucl. Phys. {\bf 568}, 447, 2000

\bibitem{4} Y. K. E. Cheung and M. Krough, Nucl. Phys. {\bf B 528} 185,
1998

\bibitem{5} V. Schomerus, Report No. JHEP 9906:030, 1999, hep-th/9903205

\bibitem{6} F. Ardalan, H. Arfaei and M.M. Sheikh-Jabbari, Report No.
JHEP 9902:016, 1999, hep-th/9810072; F. Ardalan, H. Arfaei and 
M.M. Sheikh-Jabbari, hep-th/9906161

\bibitem{7} D. Bigatti and L. Sussikind, hep-th/9908056

\bibitem{8} J.M. Maldacena and J.G. Russo, hep-th/9908134

\bibitem{9} N. Seiberg and E. Witten, hep-th/9908142

\bibitem{10} M. Li and Y.S. Wu, Phys. Rev. Lett. {\bf 84}, 2084, 2000

\bibitem{11} E.S. Fradkin and A.A. Tseytlin, Phys. Lett. {\bf B 163}, 123, 1985

\bibitem{12} C.G. Callen, C. Lovelace, C.R. Nappi and S.A. Yost, Nucl. Phys.
{\bf B 288}, 525, 1987

\bibitem{13} A. Abouelsaood, C.G. Callen, C.R. Nappi and S.A. Yost, Nucl. Phys.
{\bf B 280}, 599, 1987

\bibitem{14} M. Kontsevich, q-alg/9709040.

\bibitem{15} T.B. Mitchell, C.F. Driscoll and K.S. Fine, Phys. Rev. Lett.
{\bf 71}, 1371, 1993; K.S. Fine, A.C. Case, W.G. Flynn and C.F. Driscoll,
Phys. Rev. Lett. {\bf 75}, 3277, 1995

\bibitem{16} T. Chiueh, Phys. Rev. {\bf E 57}, 4150, 1998; 
T. Chiueh, Phys. Rev. {\bf E 61}, 3823, 2000

\bibitem{17} T. Chiueh, Phys. Rev. {\bf E 49}, 1269, 1994

\bibitem{18} P.M. Ho and Y.T. Yeh, Phys. Rev. Lett. {\bf 85}, 5523, 
2000; P.M. Ho and S.P. Miao, Phys. Rev.{\bf D 64}, 126002, 2001

\bibitem{19} A.A. Tseytlin, " Born-Infeld Action, Supersymmetry and String Theory"
in the Yuri Golfand memorial volume, ed. M. Shifman, World Scientific (2000),
hep-th/9908105

\bibitem{20} L. Susskind, hep-th/0101029
\end{thebibliography}
\end{document}